\documentclass[aps,preprint,nofootinbib]{revtex4-1}
\pdfoutput=1

\usepackage{setspace}
\usepackage{booktabs}
\usepackage{graphicx}
\usepackage{amsmath}
\usepackage{amsfonts}
\usepackage{amssymb}
\usepackage{times}
\usepackage{subfigure}
\usepackage{physics}
\usepackage{tikz}
\usepackage{float}
\usepackage{multirow}
\usepackage{slashed}
\usepackage{hyperref}
\hypersetup{colorlinks=true, linkcolor=red, urlcolor=red, citecolor=red}

\begin{document}

\title{Signatures of Ultralight  Dark Matter in Neutrino Oscillation Experiments}
\preprint{FERMILAB-PUB-20-260-T}

\author{Abhish Dev}
\email{adev@umd.edu}
\affiliation{Maryland Center for Fundamental Physics, Department of Physics, University of Maryland, College Park,MD 20742-4111 USA}

\author{Pedro A.N. Machado} 
\email{pmachado@fnal.gov}
\affiliation{Theoretical Physics Department, Fermilab, P.O. Box 500, Batavia, IL 60510, USA}

\author{Pablo Mart\'{i}nez-Mirav\'{e}} 
\email{pamarmi@ific.uv.es}
\affiliation{Departament de F{\'i}sica Te{\`o}rica, Universitat de Val{\`e}ncia, 46100, Burjassot, Spain}
\affiliation{Institut de F\'{i}sica Corpuscular CSIC-Universitat de Val\`{e}ncia,  46980 Paterna, Spain}

\begin{abstract}
We study how neutrino oscillations could probe the existence of ultralight bosonic dark matter. Three distinct signatures on neutrino oscillations are identified, depending on the mass of the dark matter and the specific experimental setup. These are time modulation signals, oscillation probability distortions due to fast modulations, and fast varying matter effects. 
We provide all the necessary information to perform a bottom-up, model-independent experimental analysis to probe such scenarios. Using the future DUNE experiment as an example, we estimate its sensitivity to ultralight scalar dark matter. Our results could be easily used by any other oscillation experiment.
\end{abstract}

\maketitle

\section{Introduction}

The nature of dark matter remains one of the greatest unknowns in particle physics. Although there is plenty of data that corroborates the existence of dark matter, its nature, mass, and interactions with the standard model are widely unknown~\cite{Tanabashi:2018oca}.
One possibility is when the mass of the dark matter (DM) is much below the electronvolt scale. For these masses, the dark matter De Broglie wavelength becomes macroscopic and could even compare to the size of solar systems and galaxies. This requires the ultralight dark matter to be heavier than $10^{-22}$~eV. Due to Fermi-Dirac statistics, phase space considerations today can be used to better constrain the mass of fermionic dark matter to be above 100 eV to the keV scale (also known as the Tremaine-Gunn bound)~\cite{Tremaine:1979we, Madsen:1990pe, Madsen:1991mz, Boyarsky:2008ju}.

Ultralight bosonic fields are common and can arise as pseudo-Nambu Goldstones from the spontaneous symmetry breaking of an approximate global symmetry such as the QCD axion~\cite{Peccei:1977np, Peccei:1977hh, Weinberg:1977ma, Wilczek:1977pj}. In addition to being ubiquitous, they could also provide dynamical solutions to finetuning problems. For example, the QCD axion makes the QCD theta angle $\theta_{QCD}$ dynamical and relaxes it to zero~\cite{Vafa:1984xg}. These fields can also constitute ultralight dark matter and result in unique fuzzy dark matter phenomenology ~\cite{Preskill:1982cy, Abbott:1982af, Dine:1982ah, Hu:2000ke, Arias:2012az, Li:2013nal, Hui:2016ltb, Menci:2017nsr, Diez-Tejedor:2017ivd, Irsic:2017yje, Visinelli:2017imh, Poddar:2019zoe}, which offers possible solutions to three small-scale cosmological puzzles, namely, cusp-vs-core, missing satellite and too-big-to-fail problems~\cite{Moore:1994yx, Flores:1994gz, Navarro:1996gj, Klypin:1999uc}. All these puzzles are based on comparisons between simulations and observed cosmological data. Compared to data, simulations predict high rotation curves towards galaxy centers due to the large DM density therein, too many satellite galaxies, and too many visible dwarf galaxies due to the presence of several DM subhalos. Fuzzy dark matter, being delocalized due to its large De Broglie wavelength, would be more susceptible to tidal disruptions and lead to broader halos possibly solving the three aforementioned problems.

Regardless of the axion solution to the strong $CP$ problem or the Fuzzy DM regime, nonzero abundance of ultralight bosonic DM could lead to a large occupation number for this field, which would behave as a vacuum expectation value (VEV).
The idea is that the DM field undergoes a phase transition in the early universe and starts oscillating around its new VEV.
Hubble friction damps these oscillations, but nevertheless, they may be still lead to observable effects in late times. 
This oscillating VEV may lead to a curious phenomenon: time variation of what we define as fundamental constants.
Such effects have been searched for in many different setups,  from resonant microwave cavities to atomic clocks (for a review, see e.g. Ref.~\cite{Battaglieri:2017aum}). Here we are interested in the possibility that neutrinos provide a portal to such ultralight fields.
This is motivated by two facts.
First, neutrinos are much lighter than other fermions in the standard model, and thus even a small time-dependent VEV could lead to a relatively large impact on neutrino masses and mixings.
Second, as the mechanism of neutrino masses remains unknown, this sector may offer a more natural connection to ultralight physics beyond the standard model.
Although we use Fuzzy DM and neutrino masses as a motivation, we are indeed more concerned with general experimental signatures, and thus we adopt a \emph{bottom-up} approach in this work.
Some of these signatures were explored in previous literature~\cite{Berlin:2016woy, Capozzi:2017auw, Krnjaic:2017zlz, Brdar:2017kbt, Farzan:2018pnk, Liao:2018byh, Ge:2018uhz, Huang:2018cwo, Ge:2019tdi, Cline:2019seo, Choi:2019zxy, Farzan:2019yvo}.
Here we will provide a more detailed analysis, with special emphasis on aspects directly connected to experimental searches, and highlighting the different regimes relevant to neutrino oscillation phenomenology.
To be concrete, we will focus on the case of ultralight scalar fields and defer ultralight vector fields to future analysis.
We will use the DUNE experiment as a case study, even if our conclusions will be valid for any neutrino oscillation experiment.

\section{Theoretical guidance to experimental searches}

First, let us review the theoretical aspects of ultralight scalar fields coupled to neutrinos.
We choose to work in an effective theory framework below electroweak symmetry breaking.
The effective Lagrangian that describes the system is
\begin{equation}
    \mathcal{L}_{\rm eff}=-m_\nu\left(1+y\frac{\phi}{\Lambda}\right)\bar\nu\nu+{\rm h.c.},
\end{equation}
where flavor indices are implicit in both $m_\nu$ and $y$, $\phi$ is the ultralight scalar field, $m_\nu$ is the neutrino mass matrix and $\Lambda$ is a heavy mass scale~\footnote{We do not distinguish Dirac and Majorana neutrinos here, as the oscillation phenomenology of these cases are identical.}.
We will not worry where this effective Lagrangian arrives from and leave considerations regarding ultraviolet completions, perhaps, to future work.

The local field value can be expressed as
\begin{equation}
    \phi(x,t)\simeq\frac{\sqrt{2\rho_\phi}}{m_\phi}\sin\left[m_\phi(t-\vec{v}\cdot \vec{x})\right],
\end{equation}
where the local density $\rho_\phi$ should not exceed the local DM density $\rho_{\rm DM}=0.3~{\rm GeV}/{\rm cm}^3$, $m_\phi$ is the mass of $\phi$ and $v\sim10^{-3}c$ is the virial velocity.
Note that the space-dependent phase $m_\phi\vec{v}\cdot\vec{x}$ of the local field value is much smaller than $m_\phi t$, and thus will be neglected henceforth.
The neutrino mass matrix will therefore receive a contribution from $\phi$ given by
\begin{equation}
   \delta m_\nu =  m_\nu y\frac{\sqrt{2\rho_\phi}}{\Lambda m_\phi}\sin(m_\phi t).
\end{equation}
Without a flavor model (see e.g. Ref.~\cite{Ding:2020yen}), the Yukawa couplings $y$ can have any structure in flavor space, and thus the modulations of neutrino masses and mixing can bear any correlation.
Nevertheless, it is useful to focus on two simple scenarios: modulation of mass splittings and modulation of mixing angles.

Besides, the mass of $\phi$ defines the modulation period via
\begin{equation}
   \tau_\phi\equiv \frac{2\pi\hbar}{m_\phi}=0.41 \left(\frac{10^{-14}~{\rm eV}}{m_\phi}\right)~{\rm seconds}.
\end{equation} 
In a given experimental setup, there are three characteristic time scales: the neutrino time of flight $\tau_\nu = L/c = 3.4 (L/1000~{\rm km})~{\rm msec}$, the time between two detected events $\tau_{\rm evt}$ (the inverse of the ratio of events) and the lifetime of the experiment $\tau_{\rm exp}$. 
From these, we can identify the following three different regimes for a given experimental setup.
\begin{itemize}
   \item {\bf Time modulation ($\tau_{\rm evt}\lesssim\tau_\phi \ll\tau_{\rm exp}$).} When the period of modulation of $\phi$ is of the same order as the experiment total run time, a temporal variation of the neutrino signal may be observed. This is true for the modulation of angles and mass splittings. Experiments with large statistics and high event rates will be sensitive to time modulation periods much smaller than the lifetime of the experiments (for instance, searches for modulations in solar neutrino fluxes are sensitive to periods ranging from 10 minutes to 10 years~\cite{Yoo:2003rc, Aharmim:2005iu, Collaboration:2009qz}).
   \item {\bf Averaged distorted neutrino oscillations ($\tau_\nu\ll\tau_\phi\ll\tau_{\rm exp}$).} Even when the rate of change of neutrino oscillation parameters is too fast to be observed as a modulating signal, the time average oscillation probability may be distorted by such effects and deviate from the standard model scenario. While the averaging of a modulating mixing angle can be mapped onto standard oscillations (with some inferred value of this mixing angle), the averaging of a mass splitting can lead to distorted neutrino oscillations (DiNOs), smearing out the probability similarly to an energy resolution smearing~\cite{Krnjaic:2017zlz}. This regime covers a large range of scalar masses and can be easily searched for in oscillation experiments, as it boils down to a simple novel oscillation effect.
   \item {\bf Dynamical distorted neutrino oscillations ($\tau_\phi\sim\tau_\nu$).} As the modulating period of $\phi$ gets closer to the neutrino time of flight, the changes in oscillation parameters need to be treated at the Hamiltonian level and can be modeled by a modified matter effect~\cite{Brdar:2017kbt}. This matter potential is time-dependent, and thus it changes as the neutrino propagates towards the detector. When the variation of the matter potential is too slow compared to the neutrino time of flight, dynamical DiNOs map onto average DiNOs. In the opposite case, when the variations are too fast compared to the neutrino time of flight, they cannot be observed and thus one recovers standard oscillations.
\end{itemize}
In the next section, we will discuss the phenomenology of each regime in general terms and highlight their phenomenology in oscillation experiments in more detail.

\section{General considerations on ultralight scalars and neutrino oscillations}

Before going into a more technical discussion, we provide a few insights on the effects of ultralight scalar fields in neutrino oscillations.
To do so, we will analyze the changes in neutrino oscillation probability in each of the regimes discussed above, using the DUNE experimental setup as a case study.
For simplicity, we will consider a single parameter modulating at a time.
Modulation of mixing angles are assumed to be of the form
\begin{equation}\label{eq:angles}
   \theta_{ij}(t)=\theta_{ij}+\eta\sin(m_\phi t),
\end{equation}
where $\theta_{ij}$ represents the undistorted value of the mixing angle and $\eta$ is the amplitude of modulation of $\phi$.
Mass splittings, on the other hand, are assumed to modulate like (for $\eta\ll 1$)
\begin{equation}\label{eq:dmsq}
   \Delta m^2_{ij}(t) \equiv m_i^2(t)-m_j^2(t) \simeq \Delta m^2_{ij}\left[1 + 2\eta \sin(m_\phi t)\right],
\end{equation}
where, similarly to above, $\Delta m^2_{ij}$ represents the undistorted value of the mass splitting.

\subsection{Time modulation}\label{sec:time-modulation}
The phenomenology in the time modulation regime is very intuitive: mixing angles or mass splittings are modulating on time.
The oscillation probability now depends on time via
\begin{equation}
   P_{\alpha\beta}\equiv P(\nu_\alpha\to\nu_\beta, t) = P\left[\nu_\alpha\to\nu_\beta; \{ \theta_{ij}(t), \Delta m^2_{ij}(t)\}\right].
\end{equation}
For example, in the case of modulating angles, the oscillation probability for $\nu_\mu\to\nu_\mu$ disappearance in vacuum, in a simplified two neutrino framework, would read
\begin{align}
   P_{\mu\mu}^{\rm angle} \simeq 1-\sin^2\left[2\theta(t)\right]\sin^2\left(\frac{\Delta m^2 L}{4E}\right)=1-\sin^2\left[2\theta+2\eta\sin(m_\phi t)\right]\sin^2\left(\frac{\Delta m^2 L}{4E}\right),
\end{align}
where $L$ is the baseline (1300~km for DUNE) and $E$ is the neutrino energy.
Notice that the oscillation probability displays a time modulation via the $\sin(m_\phi t)$ term.

\begin{figure}
    \centering
    \includegraphics[width= 0.8\textwidth]{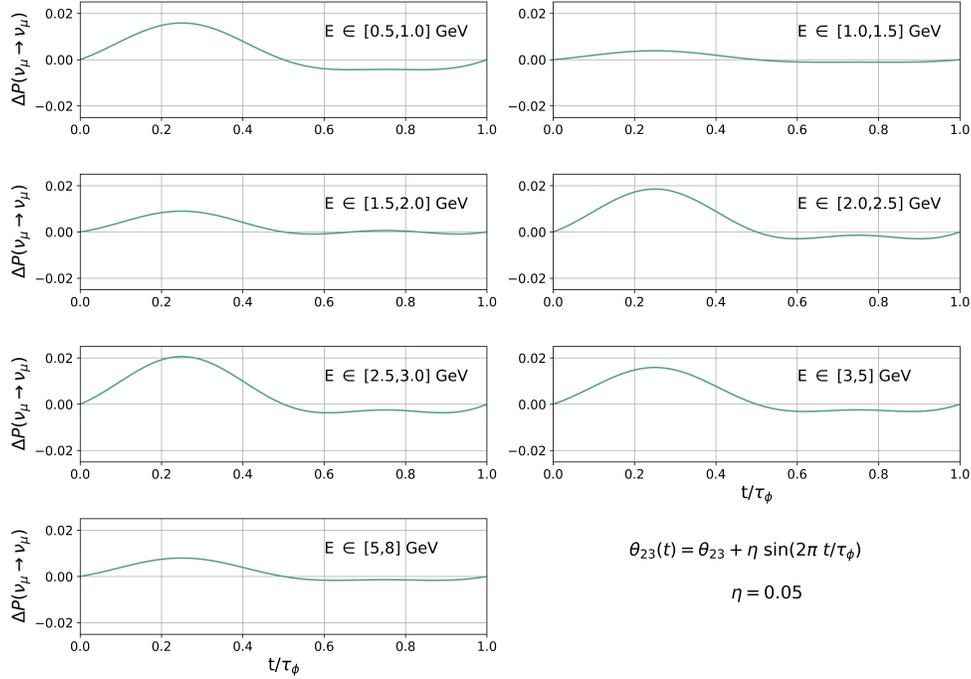}
    \caption{The difference in $\nu_\mu$ disappearance oscillation probability $\Delta P(t)\equiv P_{\mu\mu}(t)-P_{\mu\mu}(0)$ at DUNE for a modulating $\theta_{23}$ assuming $\eta=0.05$ for several energy bins. We have assumed normal mass ordering and the best fit values of oscillation parameters from Ref.~\cite{deSalas:2017kay}, namely $\Delta m^2_{31}=2.5\times 10^{-3}$~eV$^2$, and $\sin^2\theta_{23}=0.55$.}
    \label{fig:timemod-th}
\end{figure}

In Fig.~\ref{fig:timemod-th} we show change in $\nu_\mu$ disappearance oscillation probability $\Delta P(t) \equiv P_{\mu\mu}(t)-P_{\mu\mu}(0)$ at DUNE for a modulating $\theta_{23}$ assuming $\eta=0.05$ for several energy bins.
We have assumed normal mass ordering and the best fit values of oscillation parameters from \cite{deSalas:2017kay}, namely $\Delta m^2_{31}=2.5\times 10^{-3}$~eV$^2$, and $\sin^2\theta_{23}=0.55$. 
There are three important effects to be noted here. 
To  understand those, we  expand $\Delta P(t)$ for small $\eta$ as
\begin{equation}
   \Delta P(t) \simeq -2\left[\eta \sin(4\theta) \sin(m_\phi t) + 2\eta^2\cos(4\theta)\sin^2(m_\phi t)\right]  \sin^2\left(\frac{\Delta m^2 L}{4E}\right).
\end{equation}
First, the modulation phases across different energies are fully correlated, as the oscillation probability in all energy bins goes up and down at the same time.
Second, $\Delta P(t)$ depends on $\eta$ and the mixing angle itself.
And finally, when $\sin(4\theta)$ is near zero (e.g. for $\theta_{23}$ near maximal mixing), we observe a Jacobian effect that shrinks the oscillation amplitude, as the leading $\eta$ term shrinks and the $\eta^2$ term becomes dominant. 
In Fig.~\ref{fig:timemod-th}, as the modulation develops, $\sin^2 [2\theta_{23}(t)]$ goes further away from maximal, making $\Delta P(t)$ larger, and we observe the larger amplitude. 
As the phase evolves, $\sin^2 [2\theta_{23}(t)]$ gets maximal, then slightly below maximal, maximal again, and back to its original value.
This corresponds to the region to the right in the subplots ($t/\tau_\phi>0.5$).
If we had chosen to show a modulation of $\theta_{12}$ instead (e.g. in the JUNO experiment~\cite{An:2015jdp}), the change in oscillation probability would have been more symmetric.

\begin{figure}
    \centering
    \includegraphics[width = 0.8\textwidth]{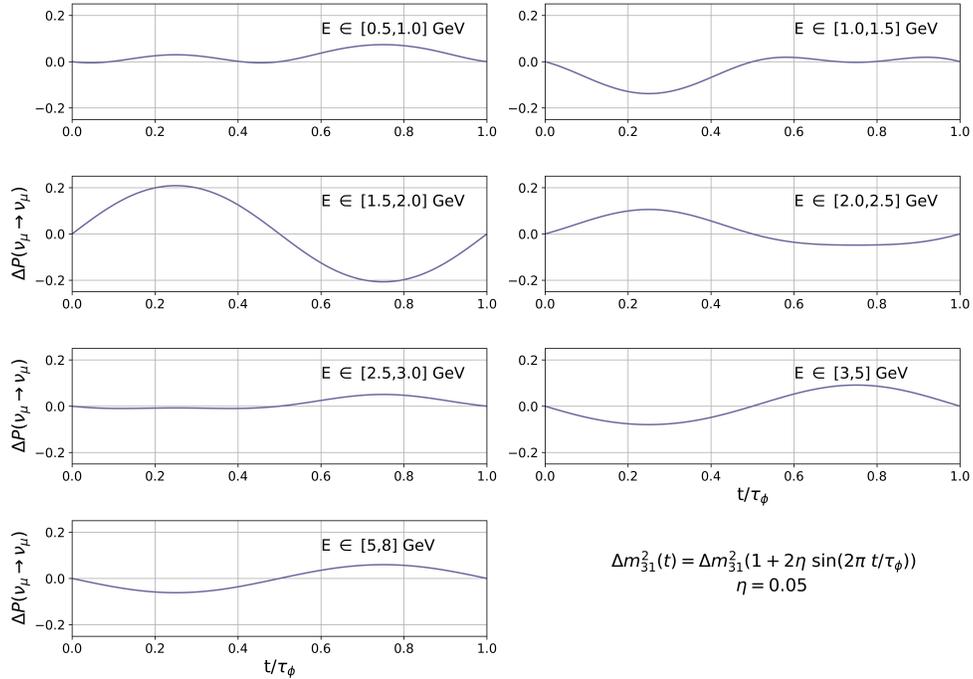}
    \caption{The difference in $\nu_\mu$ disappearance oscillation probability $\Delta P(t)\equiv P_{\mu\mu}(t)-P_{\mu\mu}(0)$  at DUNE for a modulating $\Delta m^2_{31}$ assuming $\eta=0.05$ for several energy bins. We have assumed normal mass ordering and the best fit values of oscillation parameters from Ref.~\cite{deSalas:2017kay}, namely $\Delta m^2_{31}=2.5\times 10^{-3}$~eV$^2$, and $\sin^2\theta_{23}=0.55$.}
    \label{fig:timemod-dm}
\end{figure}

If instead, the mass splittings modulate on time, the simplified time-dependent oscillation probability would read 
\begin{align}
   P_{\mu\mu}^{\rm mass} \simeq 1-\sin^2(2\theta)\sin^2\left[\frac{\Delta m^2(t) L}{4E}\right]=1-\sin^2(2\theta)\sin^2\left\{\left(\frac{\Delta m^2 L}{4E}\right) \left[1+2\eta\sin(m_\phi t)\right]\right\}. \nonumber
\end{align}
We can see that what modulates is not the amplitude of the oscillation probability, but rather the position of the minimum. In Fig. \ref{fig:timemod-dm}, we show again the change in $\nu_\mu$ disappearance oscillation probability $\Delta P(t) \equiv P_{\mu\mu}(t)-P_{\mu\mu}(0)$ at DUNE, but now for a modulating $\Delta m^2_{31}$ assuming $\eta=0.05$ for several energy bins. 
Again, we have assumed normal mass ordering and the best fit values of oscillation parameters from \cite{deSalas:2017kay}, namely $\Delta m^2_{31}=2.5\times 10^{-3}$~eV$^2$, and $\sin^2\theta_{23}=0.55$. 
The most distinctive features, in this case, are the correlations and anti-correlations across multiple energies.
This is easy to understand, as when the minimum or other regions of small probability enter/leave a given energy bin, the average probability lowers/raises in that bin.
Quantitatively, we can see these correlations in energy by expanding $\Delta P(t)$ around $\eta\to0$,
\begin{equation}
  \Delta P(t)\simeq  -2\sin^2(2\theta)\sin(m_\phi t)  \left[ \eta \xi\sin(2\xi) + 2 \eta^2 \xi^2  \cos(2\xi)\sin(m_\phi t)\right],
\end{equation}
where we have defined $\xi\equiv\Delta m^2 L/4E$.
Note that this expansion is valid  for $\eta \xi \ll 1$.
These correlations are very specific for a given experimental baseline and energy binning, and one would expect it to be useful in distinguishing ultralight scalar phenomenology from other time-dependent effects like Lorentz symmetry violation (see e.g. Ref.~\cite{Kostelecky:2011gq}).
Besides, due to the moving of the oscillation minimum, the change in oscillation probability at energies near the minimum can be enhanced.
In this example, $\eta=0.05$ can lead to $\Delta P (t)\simeq 0.2$.

A search for time-dependent frequencies in the data would seem suitable to probe this scenario, for modulations of either mixing angles or mass splittings.
Such studies have already been performed by e.g. SNO, Super-Kamiokande, and Daya Bay collaborations~\cite{Yoo:2003rc, Aharmim:2005iu, Collaboration:2009qz, Adey:2018qsd}.
In the next section, we will provide a detailed time-dependent analysis of mock experimental data (we use the DUNE set up as an example), employing the Lomb-Scargle method, to estimate sensitivity to ultralight scalars.

\subsection{Average Distorted Neutrino Oscillations}\label{sec:average-dino}
The second regime we study here is dubbed average distorted neutrino oscillations or average DiNOs for short.
In this regime, the modulation of mixing angles or mass splittings is too fast to be observed, but an averaging effect on oscillation probabilities remains.
To see how this comes about, imagine that the modulating period of the ultralight scalar field is much shorter than the experimental data collecting time, but still much longer than the neutrino time-of-flight (such that the oscillation parameters are essentially constant for each neutrino event).
The average oscillation probability would be given by~\cite{Krnjaic:2017zlz}
\begin{align}
   \langle P_{\alpha\beta}\rangle=\frac{1}{\tau_\phi}\int_0^{\tau_\phi} dt \, P_{\alpha\beta}(t),
\end{align}
where $\tau_\phi$ is the period of the ultralight scalar field.
First, we focus on the averaging of mass splittings, which will prove to be more interesting than averaging of angles.
For the vacuum, 2-flavor, muon neutrino disappearance oscillation, the averaging is
\begin{align}\label{eq:average-prob}
   \langle P^{\rm mass}_{\alpha\beta}\rangle & =\frac{1}{\tau_\phi}\int_0^{\tau_\phi} dt \left\{ 1-\sin^2(2\theta) \sin^2\left[\left(\frac{\Delta m^2 L}{4E}\right) \left[1+2\eta\sin(m_\phi t)\right]\right]\right\}\\
        & \simeq 1-\sin^2(2\theta)\left\{\sin^2\left(\frac{\Delta m^2 L}{4E}\right)+2\eta^2 \left(\frac{\Delta m^2 L}{4E}\right)^2\cos\left(\frac{\Delta m^2 L}{2E}\right)\right\},
\end{align}
\begin{figure}
    \centering
    \includegraphics[width = 0.7\textwidth]{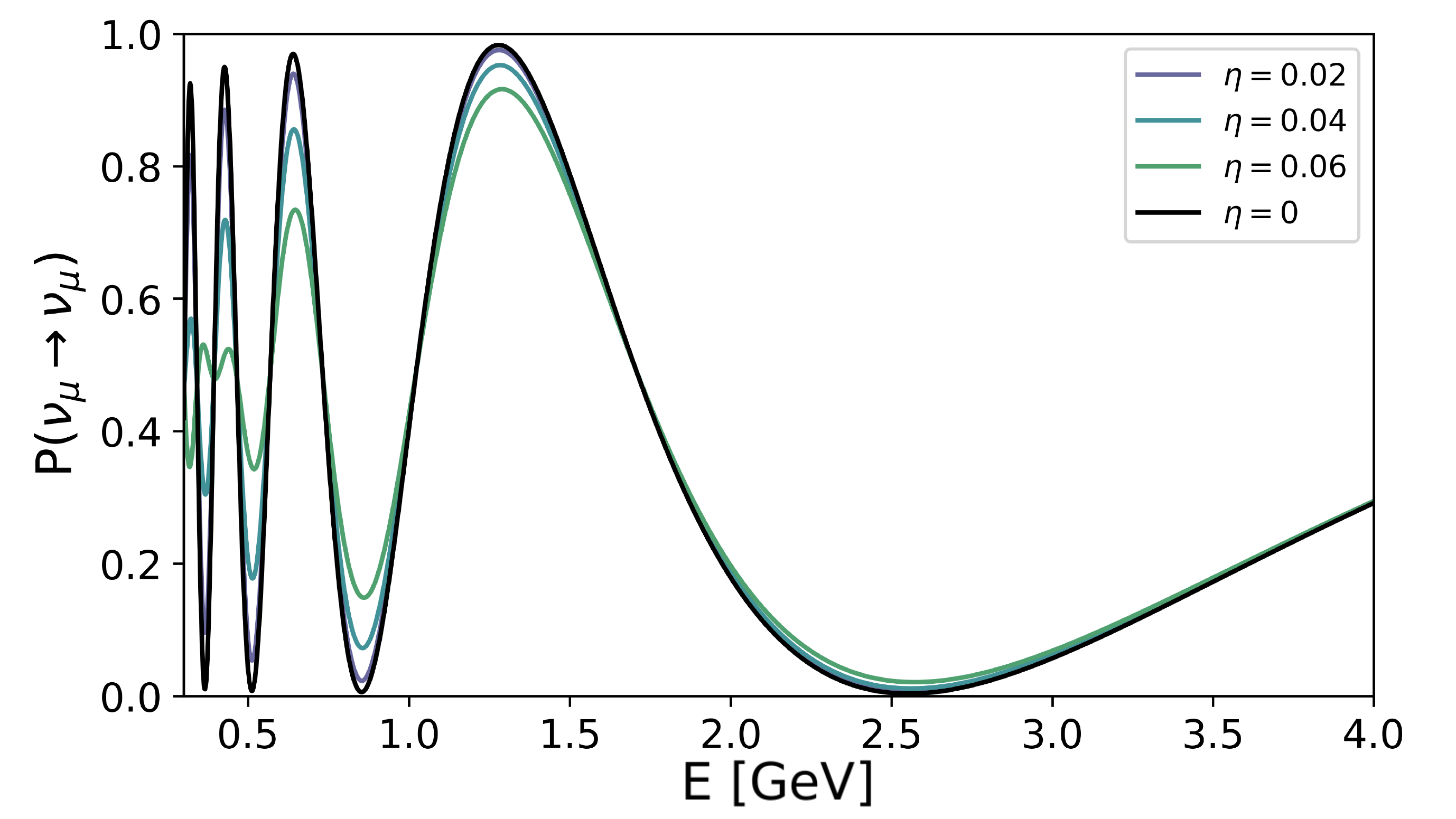}
    \caption{Averaged $\nu_\mu$ disappearance probability at DUNE for a modulating $\Delta m^2_{31}$ for three different values of $\eta$ and in the absence of DiNOs ($\eta$= 0).}
    \label{fig:av-dm}
\end{figure}
where the last expression was expanded to order $(\eta \Delta m^2 L/4E)^2$, which is typically fine for the first oscillation minimum and $\eta<0.05$. 
As shown in Fig.~\ref{fig:av-dm} for oscillation parameters from Ref.~\cite{deSalas:2017kay} and different values of $\eta$, the effect of mass splitting averaging is a smearing in the oscillation probability, similar to the effect of a finite energy resolution.
Therefore, one would expect experiments like DUNE, KamLAND and JUNO to be ideal to probe such scenarios, as all of them have good energy resolution and are endowed with broad band beams that allow for the observation of the full shape of the oscillation probability.

Regarding the averaging of modulating angles, it is easy to see that this effect simply maps into standard oscillation probability with different parameters.
The average mixing would be~\cite{Krnjaic:2017zlz}
\begin{align}
   \frac{1}{\tau_\phi}\int_0^{\tau_\phi} dt \, \sin^2\left[2\theta+2\eta\sin(m_\phi t)\right] = \frac{1}{2}\left[1-J_0(4\eta)\cos(4\theta)\right]\simeq \sin^2(2\theta)(1-4\eta^2)+2\eta^2,
\end{align}
where $J_0$ is a Bessel function of the first kind and the righthand side was expanded to second order in $\eta$.
As one expects, the effect of averaging is pushing apparent mixing angles away from zero or maximal mixing.
Experimental sensitivity to average DiNO effects on mixing angles depends not only on the precision with which the experiment can determine the mixing angles but also on the value of the measured angle itself.
Because of that, we will focus on the modulation of mass splittings in the case of average DiNO.

\subsection{Dynamical Distorted Neutrino Oscillations}\label{sec:dynamical-dinos}
The last regime is the dynamical DiNO, in which the effect of modulation of the neutrino mass matrix in oscillations needs to be taken into account at the Hamiltonian level.
The matter potential induced by the ultralight scalar field can be written as~\cite{Brdar:2017kbt} 
\begin{equation}
  V_\phi(t) = \frac{1}{2E}\left[(Y m_\nu + m_\nu Y)\phi(t)+Y^2\phi^2(t)\right],
\end{equation}
where we have defined the coupling matrix $Y\equiv m_\nu y/\Lambda$. 
Then the Hamiltonian that drives the evolution of the system is given by
\begin{equation}
  H(t) = H_{\rm vac} + V_{\rm matter}+ V_\phi(t),
\end{equation}
where $H_{\rm vac}$ is the vacuum Hamiltonian, that is, $\textrm{diag}(0,\,\Delta m^2_{21},\, \Delta m^2_{31})/2E$ in the mass basis, $V_{\rm matter}$ is the usual matter potential, $V_{\rm matter} = U^\dagger \textrm{diag}(V_{\rm cc},\,0,\, 0)U$ in the mass basis, and $U$ is the PMNS matrix.
Since $\phi$ evolves in time, the full Hamiltonian also depends on time.

Here we propose a simplification of this treatment which is valid when considering a modulation effect either solely in the mixing angles  or solely in the mass splittings.
If the modulation of $\phi$ affects solely the mass splittings, we can work in the mass basis (denoted by a ``0'' subscript) and write
\begin{equation}
     H_0(t) \equiv \frac{1}{2E}
\left(\begin{array}{ccc}
 0 & 0  &0   \\
 0 & \Delta m^2_{21}(t)  & 0  \\
 0 & 0  & \Delta m^2_{31}(t)
\end{array}\right)
   +U^\dagger \left(\begin{array}{ccc}
 V_{\rm cc} & 0  & 0  \\
 0 & 0  & 0  \\
 0 & 0  & 0
\end{array}\right)U
\end{equation}
where the time dependent mass splittings are given in Eq.~\eqref{eq:dmsq}.
We write the neutrino state in the mass basis as $\nu = U^\dagger \nu_{\rm fl}$ (the ``fl'' subscript denotes flavor). 

The evolution of this state is given by   $\dot\nu = i H_0(t) \nu$, which has to be solved numerically and leads to an ``instantaneous'' oscillation probability from flavor $\alpha$ to $\beta$, namely, $P_{\alpha\beta}(t_0,L)$.
Numerically, the propagation of the neutrino from the source to the detector can be implemented by dividing the path into $N$ layers of thickness $\Delta L=L/N$ where the oscillation parameters do not change substantially in each layer. 
The oscillation probability from flavor $\alpha$ to flavor $\beta$ is given by
\begin{equation}
    P_{\alpha\beta} (t_0, L) = |\bra{\nu_\alpha} U \left\{\prod_{n=1}^N \exp\left[i H_0(t_n) \Delta L\right]\right\}U^\dagger\ket{\nu_\beta}|^2~~~~~~~~~(\textrm{for } \Delta m^2 \textrm{  modulation}),
\end{equation}
where $U^\dagger\ket{\nu_{\alpha}}$ is a flavor state $\ket{\nu_{\alpha}}$ written in the mass basis, and $t_n\equiv t_0+n\Delta L$. 
Note that due to the presence of the MSW term, the Hamiltonian in different layers do not commute with each other. Throughout the duration of the experiment, all possible initial phases will be scanned randomly. 
Therefore, the observed oscillation probability is the time average of $P_{\alpha \beta}(t_0,L)$, namely
\begin{equation}\label{eq:obs-prob}
  \langle P_{\alpha\beta}(L)\rangle = \frac{1}{\tau_\phi}\int_0^{\tau_\phi}dt_0 P_{\alpha \beta}(t_0,L),
\end{equation}
where again, $\tau_\phi=2\pi/m_\phi$ is the period of oscillation of $\phi$.


For the case of modulating mixing angle, it is more convenient to write the Hamiltonian in the flavor basis (again denoted by the subscript ``fl'') and write the full Hamiltonian as
\begin{equation}
     H_{\rm fl}(t) \equiv \frac{1}{2E}U(t)
\left(\begin{array}{ccc}
 0 & 0  &0   \\
 0 & \Delta m^2_{21}  & 0  \\
 0 & 0  & \Delta m^2_{31}
\end{array}\right)U^\dagger(t)
   + \left(\begin{array}{ccc}
 V_{\rm cc} & 0  & 0  \\
 0 & 0  & 0  \\
 0 & 0  & 0
\end{array}\right),
\end{equation}
where the PMNS matrix is now changing in time.
The evolution of a neutrino of definite flavor $\nu_{\rm fl}$ is $\dot\nu_{\rm fl} = i H_{\rm fl}(t) \nu_{\rm fl}$ and the observed oscillation probability is given by Eq.~\eqref{eq:obs-prob}.
Following the previous recipe, the oscillation probability can be calculated with
\begin{equation}
    P_{\alpha \beta} (t_0, L) = |\bra{\nu_\alpha} \left\{\prod_{n=1}^N \exp\left[i H_{\rm fl}(t_n) \Delta L\right]\right\}\ket{\nu_\beta}|^2~~~~~~~~~(\textrm{for } \theta_{ij} \textrm{  modulation}).
\end{equation}
Note that we do not need to rotate the initial flavor neutrino by the PMNS matrix as we decided to work here in the more appropriate flavor basis. 
The observed oscillation probability is again given by the average over $t_0$, as in Eq.~\eqref{eq:obs-prob}.

\begin{figure}
    \includegraphics[width = 0.47\textwidth]{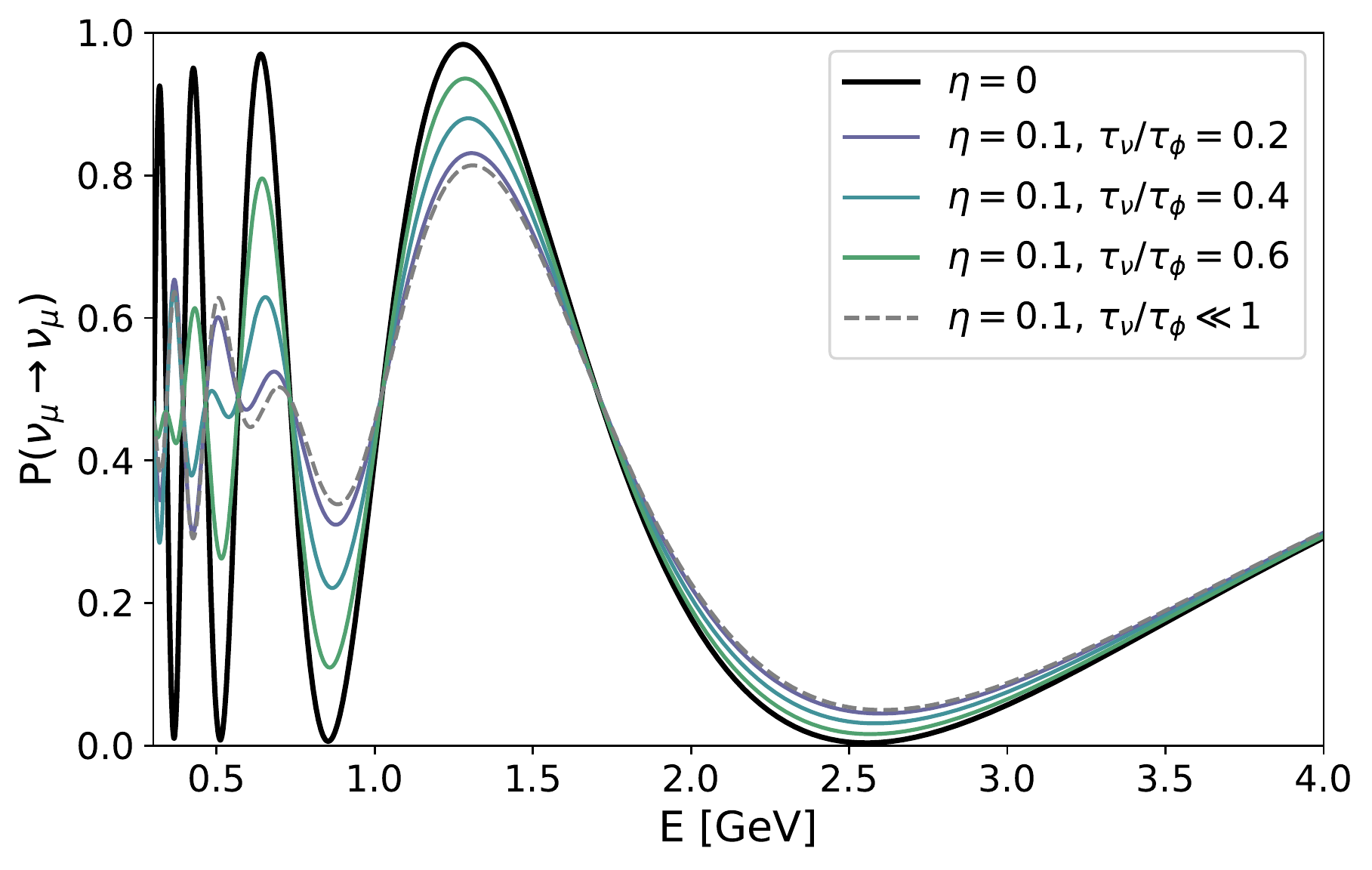}~
    \includegraphics[width = 0.47 \textwidth]{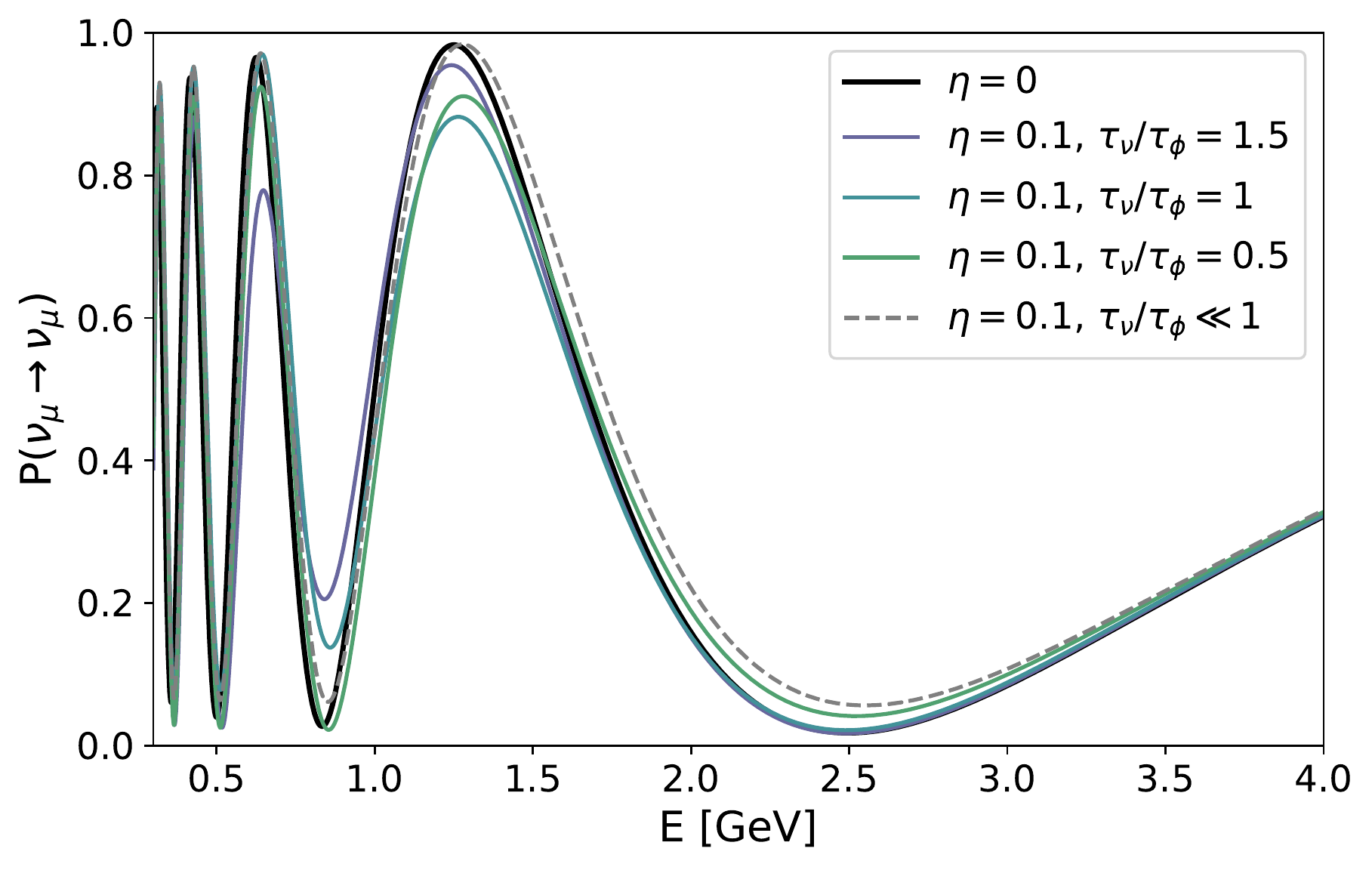}
    \caption{The $\nu_\mu$ disappearance probability at DUNE for a modulating $\Delta m^2_{31}$ (left) and $\theta_{23}$ (right) in the dynamical regime, for different values of the ratio between the period of the modulation $\tau_\phi$ and the neutrino time of flight $\tau_{\nu}$.  }
    \label{fig:dynprob}
\end{figure}

We can see the effect of dynamical DiNOs in Fig.~\ref{fig:dynprob} for modulations of $\Delta m^2_{31}$ (left) and $\theta_{23}$ (right) for oscillation parameters from Ref.~\cite{deSalas:2017kay} and a modulation amplitude of $\eta=0.1$ and $L=1300$~km.
In the left panel, it is clear to see that as $\tau_\phi/\tau_\nu$ gets larger, the effect shrinks, as the changes in the matter effect become too fast to affect neutrino oscillations.
Besides, as $\tau_\phi/\tau_\nu$ gets smaller, the effect of dynamical DiNOs asymptotes to that of average DiNOs, as one would expect. 
For mixing angle modulations (right panel), the Jacobian effect discussed in Sec.~\ref{sec:time-modulation} suppresses the impact of modulating $\theta_{23}$ in DUNE.
Notice also that there is a small displacement of the minima and maxima of oscillations.
This is simply due to the fact that the effective mass splitting measured in long baseline $\nu_\mu$ disappearance is not exactly $\Delta m^2_{31}$, but rather a function of atmospheric splittings and mixings, typically dubbed $\Delta m^2_{\mu\mu}$~\cite{Nunokawa:2005nx}.
In this plot, we have chosen a fixed value of $\Delta m^2_{31}$ to obtain the curves.
This displacement could simply be mapped into a different value of the atmospheric mass splitting. 

\section{Case study: DUNE phenomenology}

In this section, we will perform a case study of how the oscillation experiments can probe the ultralight scalars that we introduced in the previous sections. We will analyze the DUNE sensitivity to the aforementioned three regimes.
For all analyses performed here, we followed the simulation of Ref.~\cite{Alion:2016uaj} using the GLoBES software~\cite{Huber:2004ka, Huber:2007ji}
We have assumed the DUNE experiment to have a 1.07~MW beam, four far detectors with a total mass of 40~kton, and a total run time of 7 years, equally divided between neutrino and antineutrino mode.
We took the matter density to be constant $\rho=2.8~{\rm g}/{\rm cm}^{3}$.
The neutrino spectrum at DUNE spans energies roughly from 1-5 GeV, with a peak at around 3 GeV.
From Ref.~\cite{Alion:2016uaj}, the main systematic uncertainties are related to the beam and cross-section modeling.
As it is important for some of the results that will follow, we call the attention that the reconstructed neutrino energy resolution from Ref.~\cite{Alion:2016uaj} is approximately 16\% at 3~GeV.
Finally, in general terms, the ultralight phenomenology is better probed by the $\nu_\mu$  and $\bar\nu_\mu$ disappearance channels, as they bring in the largest statistics.

In our numerical simulations, the best fit oscillation parameters were taken from Ref.~\cite{deSalas:2017kay} ($\Delta m^2_{21}=7.5\times 10^{-5}$~eV$^2$, $\Delta m^2_{31}=2.5\times 10^{-3}$~eV$^2$, $\sin^2\theta_{12}=0.32$, $\sin^2\theta_{13}=0.0216$, $\sin^2\theta_{23}=0.55$, $\delta=1.2\pi$) and normal mass ordering was assumed for concreteness. 
We have used current priors for the oscillation parameters that were marginalized over in the analysis performed~\footnote{We took the largest one-sided $1\sigma$ error for each oscillation parameter from Ref.~\cite{deSalas:2017kay}).}.

\subsection{Time modulation phenomenology at DUNE}

The signature of the time modulation regime is the presence of a periodic signal in the oscillated neutrino spectrum at the far detector.
As can be seen in Eqs.~\eqref{eq:angles} and \eqref{eq:dmsq}, the period of modulation is given by the mass of the scalar field, and thus a positive signal of time modulation at DUNE would provide a measurement of the mass of this field.

In general, the search for periodic signals in data sets can be performed using the Lomb-Scargle periodogram~\cite{Lomb:1976wy, Scargle:1982bw}, which is an extension of the classical periodogram for unevenly separated data (see Ref.~\cite{VanderPlas_2018} for a pedagogical review).
Such searches are not new in the context of neutrino physics. 
For instance, many analyses of periodicities in the solar neutrino flux have been performed in SNO~\cite{Aharmim:2005iu, Ranucci:2006rz} and SuperKamiokande~\cite{Ranucci:2005ep}, as well as for other time-varying signals in the context of Lorentz and CPT violation in Daya Bay~\cite{Adey:2018qsd}. 

To evaluate the statistical significance of a modulation in a data set, we first define the Lomb-Scargle (LS) power for a frequency $\omega$ as 
\begin{equation}
 \begin{split}
    P_{LS}(\omega) = 
\frac{1}{2}\Bigg\{ \bigg( \sum_{n} g_n \cos [2\pi \omega (t_n -\tau)]\bigg)^2 \bigg/ \sum _n \cos^2 [2\pi \omega (t_n -\tau)] + \\
\bigg( \sum_{n} g_n \sin [2\pi \omega (t_n -\tau)]\bigg)^2  \bigg/ \sum _n \sin^2 [2\pi \omega (t_n -\tau)]   \Bigg\}
\end{split}   
\end{equation}
where $g_n = g(t_n)$ is the signal at the time of the measurement $t_n$ and $\tau$ is defined by solving $\tan (4\pi \omega \tau) = \sum_n \sin(4 \pi \omega t_n) / \cos(4\pi \omega t_n)$.

The significance of a given LS power can be quantified with the False Alarm Probability Test (FAP), which is a measure of how likely it is that a data set with no signal would give rise to a peak of the same magnitude as a consequence of spurious background noise. To estimate the FAP we followed the Baluev approach~\cite{Baluev:2007su}, which provides an upper limit to the value obtained through more sophisticated approaches based on a Bootstrap Method. A more detailed discussion can be found in Ref.~\cite{VanderPlas_2018}.

In our analysis, we considered larger than nominal energy bins, which allows us to extend the sensitivity to smaller periods. 
In order to cover all the parameter space presented, different time binning of the events has been explored. 
In an experimental setup, different choices of the time bins can be explored \textit{a posteriori}, ensuring that the whole accessible parameter space is covered.
As discussed in Fig.~\ref{fig:timemod-dm}, time modulation effects from ultralight scalars can lead to correlated (and anti-correlated) modulations in different energy bins, depending on which oscillation parameter is modulating.
In the case of a positive modulation signal, such correlations can be exploited to further constrain the model.
Different data sets generated for different values of modulation amplitude and frequency have been tested with the LS periodogram and Baluev's approach to the FAP test. 

\begin{figure}[t!]
    \centering
    \includegraphics[width = 0.375\textwidth]{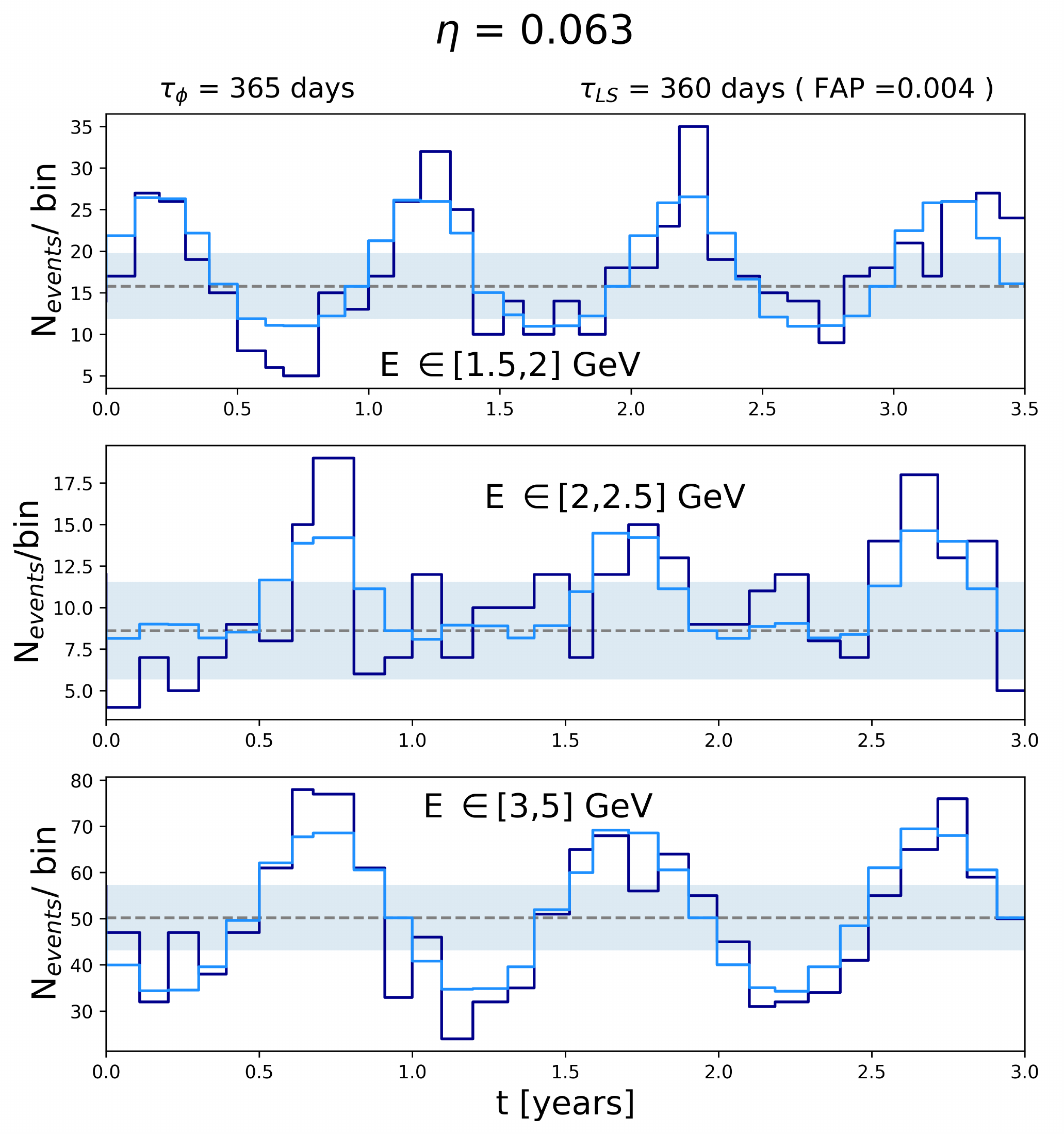}
    \hspace{1cm}
    \includegraphics[width = 0.545\textwidth]{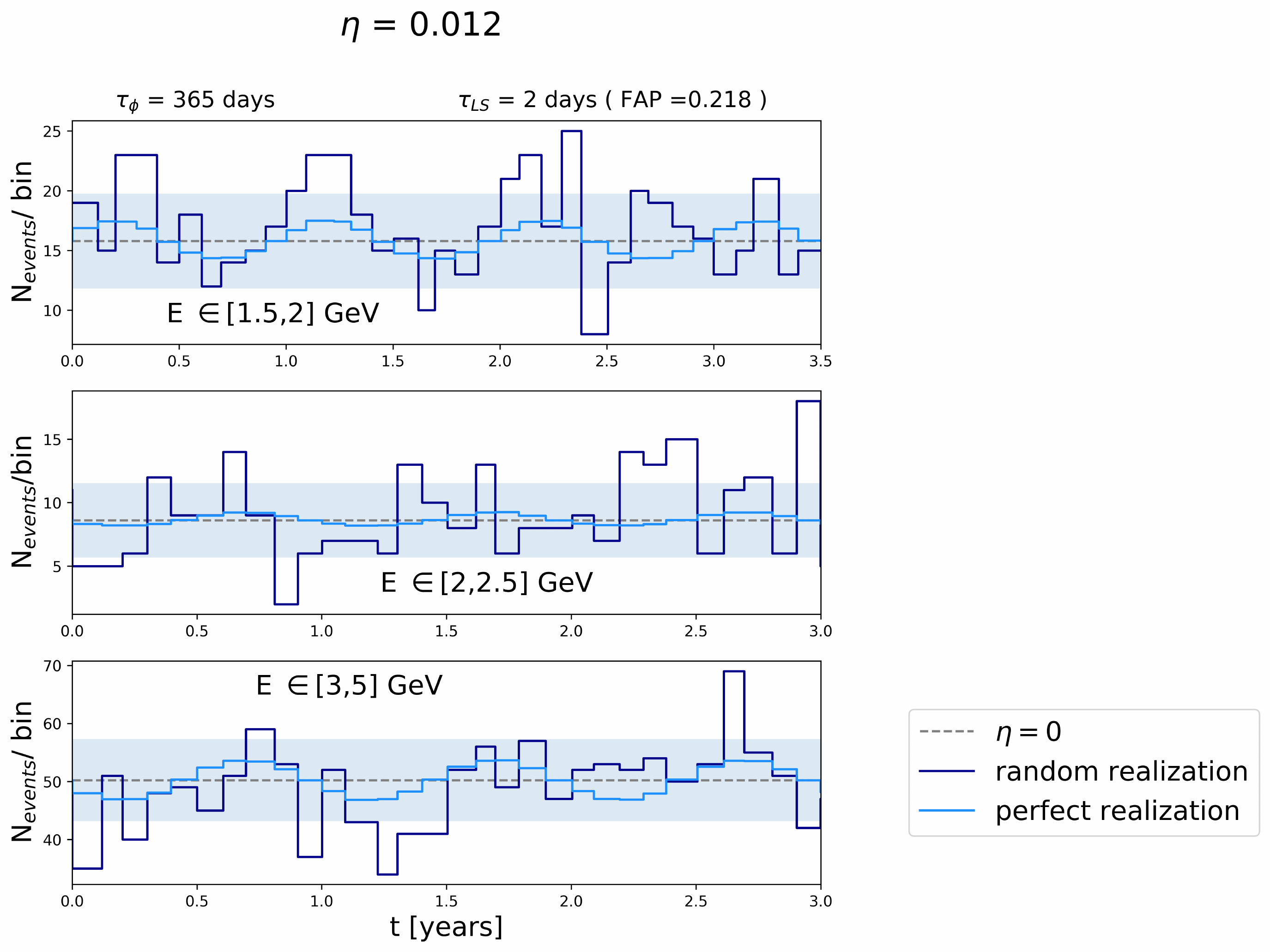}
    \caption{Expected number of events in different time bins at three different energy bins for a particular realisation of DUNE. Gray dashed lines represent the number of events expected with $\eta = 0$ and the shadowed region corresponds to the $1\sigma$ statistical uncertainty. Light blue lines are the expected  signal with non-zero $\eta$ and $m_\phi$, while dark blue lines represent a random realisation of the signal which includes statistical fluctuations.
    Left panel corresponds to a set of parameters $(\eta, m_\phi)$ in the 90$\%$ CL region above shown. Right panel corresponds to a point in the $\eta-m_\phi$ plane which LS method would not identify the frequency correctly. 
    \label{fig:allowed}}
\end{figure}

A typical realization of the signal in DUNE, for modulation of mass splittings, is presented in Fig.~\ref{fig:allowed} for illustrative purposes. 
Two modulation amplitudes are shown, $\eta=0.063$ and $\eta=0.012$.
As we will see later, DUNE is expected to be sensitive to the larger but not to the smaller.
The expected number of events is presented for three different energy bins assuming 3 years running time in the neutrino mode~\footnote{Although we show only 3 energy bins, we use the entire energy range available in our simulations.}.   
The original period of the data ($\tau_\phi$) and the one determined with the LS method ($\tau_{LS}$) are presented together with the FAP score.

\begin{figure}
    \centering
    \includegraphics[width = 0.8\textwidth]{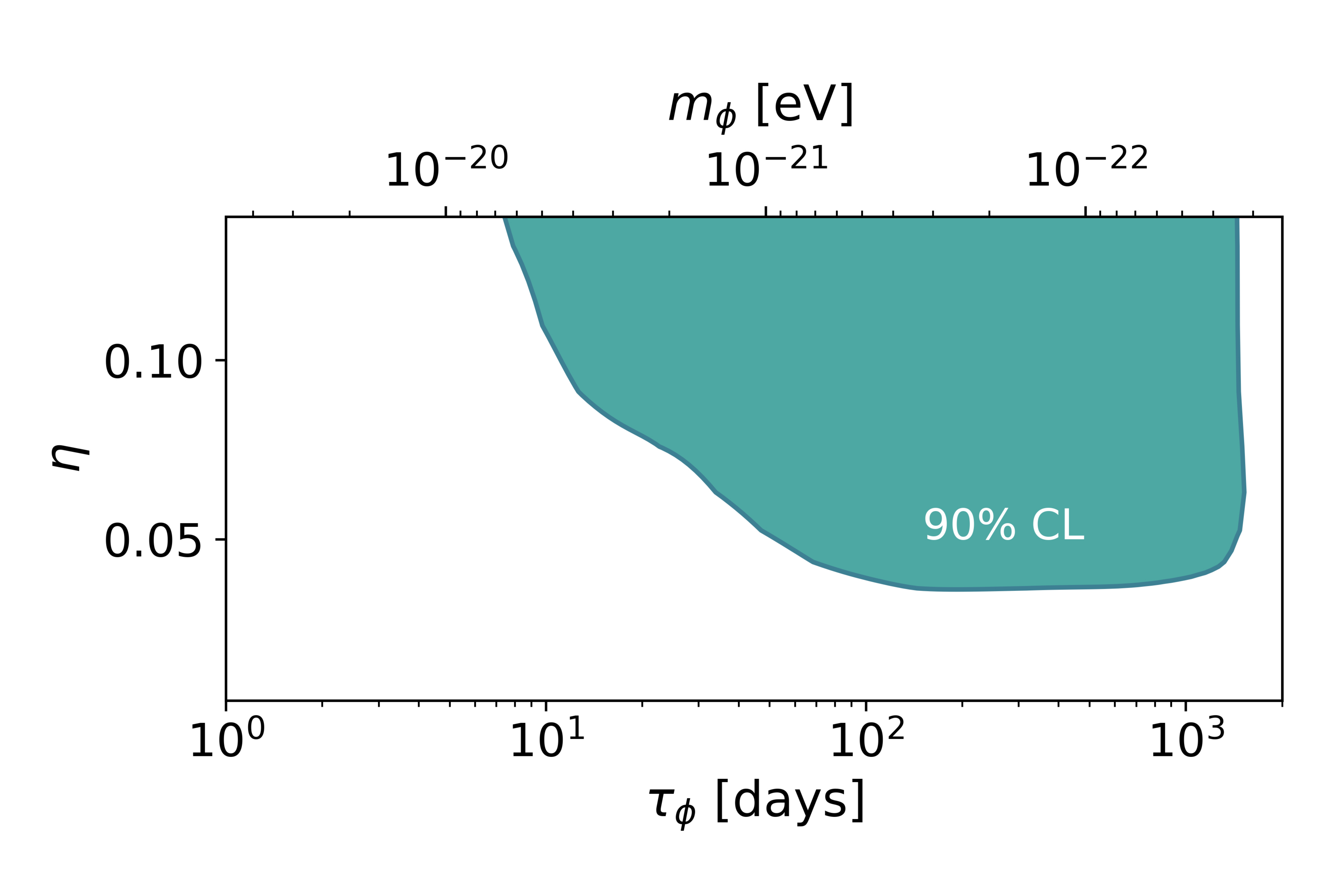}
    \caption{Region in the modulation amplitude $\eta$ versus period $\tau_\phi$ plane for which the analysis performed using the Lomb-Scargle periodogram would be capable to state at a 90$\%$ C.L. that the periodic signal found in 7 years of DUNE data would not be spurious. Upper x-axis presents the corresponding range for the mass of the ultralight scalar, $m_\phi$.}
    \label{fig:LS}
\end{figure}

In Fig.~\ref{fig:LS}, we show the DUNE sensitivity to modulations of the mass splitting $\Delta m^2_{31}(t)$ as discussed in Eq.~\eqref{eq:dmsq}.
We present the parameter space for which the Lomb-Scargle method would identify a frequency in the data set such that one would be able to state that there is a 90$\%$ probability that the periodic signal found in data is not due to random noise. 
It is important to point out that this is different from stating that there is a periodic signal with a given frequency in the data set.

The smallest period (largest frequency) to which DUNE is expected to be sensitive corresponds to the case in which the scalar oscillation period is about a few days. 
Sensitivity at large frequencies requires smaller time bins, leading to fewer events per bin which are more prone to statistical fluctuation. 
Consequently, the sensitivity to the amplitude decreases when looking for shorter periods. 
The largest period is determined by the running time of the experiment and is expected to be a couple of years.  This means that DUNE would be sensitive to the range of scalar masses  $5\times 10^{-23}\lesssim m_\phi \lesssim 10^{-20}$~eV.
Moreover, the maximum sensitivity on the amplitude of the modulation $\eta$ is about 4\%.

We do not show the DUNE sensitivity to modulations of the mixing angle $\theta_{23}$. 
Due to the Jacobian effect discussed in Sec.~\ref{sec:time-modulation}, the values of $\eta$ necessary to yield an observable effect at DUNE would need to be $\mathcal{O}(1)$. 
Therefore we see the mass splitting modulation search as a more promising way of detecting the presence of ultralight bosonic dark matter.
Nevertheless, the same Lomb-Scargle technique described above could be used to probe the modulations of mixing angles.

This analysis is not free from systematic uncertainties.
Deadtime intervals of any neutrino detector may span from months to milliseconds.
These would constitute an important source of systematic errors when analyzing time modulations. 
In particular, quasiperiodic deadtimes such as scheduled breaks of runs or maintenance and calibration operations could lead to peaks in the Lomb-Scargle power spectrum. 
It is possible to deal with them by estimating the window function, see Ref.~\cite{VanderPlas_2018}. 
This method allows us to identify the main features of the structure of the window. 
The existence of a certain periodicity in data taking would induce peaks both in the Lomb-Scargle power spectrum and in the window power spectrum and consequently, one can identify the corresponding frequencies as related to the experiment and not to the physical phenomenon under study. 

Beam unrelated backgrounds, like cosmic rays and atmospheric neutrino, can also exhibit time modulation, but these are typically negligible in beam neutrino experiments.
Moreover, beam performance can change over time.
One would expect near-to-far ratios to be less sensitive to these variations.
Nevertheless, the observed neutrino beam is not the same in the near and far detectors (due to different geometry), and oscillations may further enhance this difference.
Thus, experiments should take those systematics into consideration when performing a search for the time dependence of the neutrino signal.
In our results, we have not considered those systematics, though we believe they would not degrade the overall sensitivity by much.

\subsection{Average distorted neutrino oscillations at DUNE}

The second regime we will study in detail in DUNE is the case of average distorted neutrino oscillations (average DiNOs). 
In this case, modulations of mass splittings are too fast to be observed as a time modulation signal, but the averaging of the modulation still imprint observable effects in the oscillation probability.
As discussed in Sec.~\ref{sec:average-dino}, the modulation of mass splittings leads to more interesting phenomenology than the modulation of mixing angles.
If $\Delta m^2 _{31}$  varies in time, the maxima and minima are displaced periodically. 
For a very fast modulation, such displacement manifests as a non-trivial averaging and has to be carefully studied in order to disentangle it from the distortion caused by the finite energy resolution~\cite{Krnjaic:2017zlz}. 
Consequently, the searches here presented would benefit from improvements in the energy resolution, as the one proposed in Ref.~\cite{Friedland:2018vry}.
As a general rule of thumb, the new physics effect here is just a modification of the oscillation probability, so all systematic uncertainties associated to an oscillation search would be relevant in this case.
We include those systematics using the simulation from Ref.~\cite{Alion:2016uaj}.

\begin{figure}
    \centering
    \includegraphics[width = 0.58\textwidth]{figures/chi2-dmav.pdf}~    
    \includegraphics[width = 0.4\textwidth]{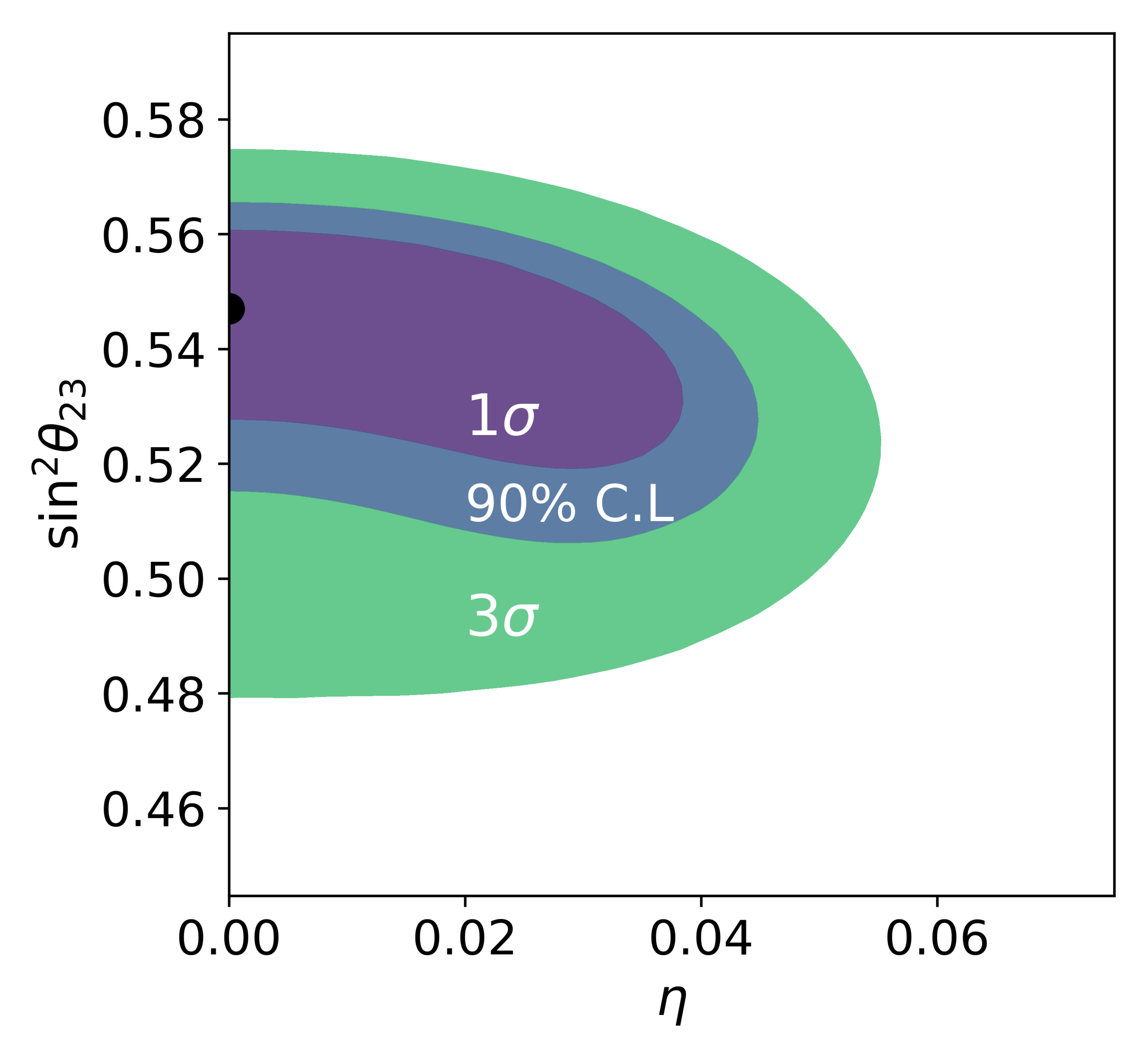}
    \caption{Left: $\chi^2$ profile as a function of $\eta$ and confidence levels to discriminate with respect to the standard 3$\nu$ picture. Right: allowed region in the plane $(\eta, \, \sin^2\theta_{23})$ in DUNE. Best fit values chosen for mock data are marked by a black circle (standard oscillations assumed as true hypothesis).}
    \label{chi-av}
\end{figure}

The average oscillation probability in Eq.~\eqref{eq:average-prob}, necessary to estimate DUNE sensitivity to this scenario, was calculated by numerically averaging the analytic approximations for neutrinos oscillations in the presence of matter from Ref.~\cite{Denton:2016wmg}. 
The DUNE sensitivity  to fast modulations of the mass splitting is shown in Fig.~\ref{chi-av}.
The range of periods that DUNE is sensitive to, in this regime, is approximately from a year ($\tau_\phi\ll\tau_{\rm exp}\sim$~10 years) to tens of milliseconds ($\tau_\phi\gg\tau_\nu\simeq$~4.3~msec).
This translates into a very large range of masses $2\times 10^{-23} \lesssim m_\phi\lesssim 3\times 10^{-14}$~eV.
On the flip side, if a signal of average DiNOs is observed at DUNE, it would be very challenging to pinpoint the exact mass of $\phi$, as its effects has been averaged out.
As can be seen from the left panel of Fig.~\ref{chi-av}, DUNE is expected to be sensitive to roughly 4\% modulation amplitudes of the mass splitting at  90\% C.L.
In estimating DUNE's sensitivity, we have marginalized over all oscillation parameters.
Note that the energy resolution is crucial here, and we expect improvements in energy resolution to translate into better sensitivity to average DiNOs.

Besides, one may ask what is the impact on the determination of the standard mixing parameters if an average DiNO effect is present.
We show in the right panel of Fig.~\ref{chi-av} the allowed region by DUNE in the plane $(\eta,\,\sin^22\theta_{23})$.
There is some mild degeneracy between $\theta_{23}$ and the modulation amplitude $\eta$ for small values of $\eta$. 
This is because, for small $\eta$, the minimum of the $\nu_\mu\to\nu_\mu$ oscillation probability lifts slightly thus allowing values of $\theta_{23}$ closer to $\pi/4$.
We have checked that there is no degeneracy between $\delta_{CP}$ and $\eta$.

\subsection{Dynamical distorted neutrino oscillations}
Now, we proceed to the last regime of ultralight scalar field phenomenology in neutrino experiments, that of dynamical distorted neutrino oscillations (dynamical DiNOs). 
In this regime, the time of flight of the neutrino is comparable to the magnitude of the scalar period and we have to take full account of the variation of the oscillation parameters during the journey from the source to the detector. 
As before, we consider modulations in $\theta_{23}$ and $\Delta m^2_{31}$ separately, which leads to the oscillation probabilities discussed in Sec.~\ref{sec:dynamical-dinos}.
As discussed in the previous section, the new physics effect here is again just a modification of the oscillation probability, so all systematic uncertainties associated with an oscillation search would be relevant in this case.

\begin{figure}
    \centering
    \includegraphics[width = 0.8\textwidth]{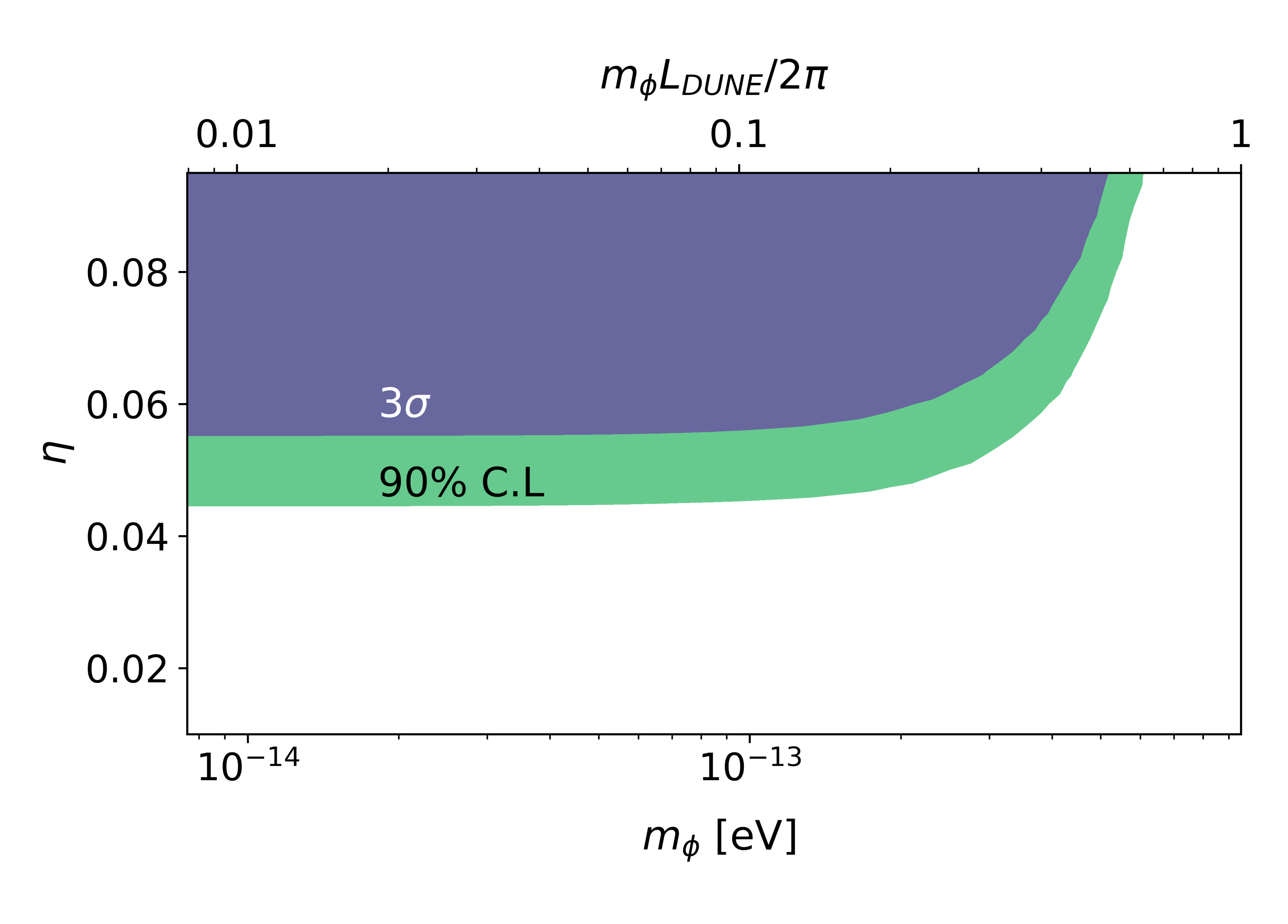}
    \caption{DUNE sensitivity in a fast modulation of the $\Delta m^2_{31}$ mass splitting with amplitude $\eta$ versus the mass of the ultralight scalar $m_\phi$ plane at 90$\%$ C.L (green) and $3\sigma$ (blue). Upper x-axis presents the corresponding range for the dimensionless quantity $m_\phi L_{DUNE} / 2\pi$.}
    \label{fig:dyn-mass}
\end{figure}

We evaluate the DUNE sensitivity to the dynamical DiNO regime for the case of mass splitting modulations and present the results in Figs.~\ref{fig:dyn-mass}.
As expected, when the modulation period is much smaller than the neutrino time-of-flight (of about 4.3~msec), the experimental sensitivity degrades as the oscillation probability tends to the standard one.
Besides, for mass splitting modulation, when the modulation period is sufficiently large, we recover the average DiNO sensitivity, see Fig.~\ref{fig:dynprob}.
In the case of mixing angle modulation, the averaging simply maps the modulating mixing angle onto another value of the mixing, and thus we do not analyze this case here.
In principle, a richer flavor structure in the coupling between ultralight scalar and neutrinos would allow concomitant modulation of mass splittings and angles with arbitrary correlations. 
For simplicity, we do not pursue this possibility in this manuscript.

\vspace{1cm}

\begin{figure}
    \centering
    \includegraphics[width = \textwidth]{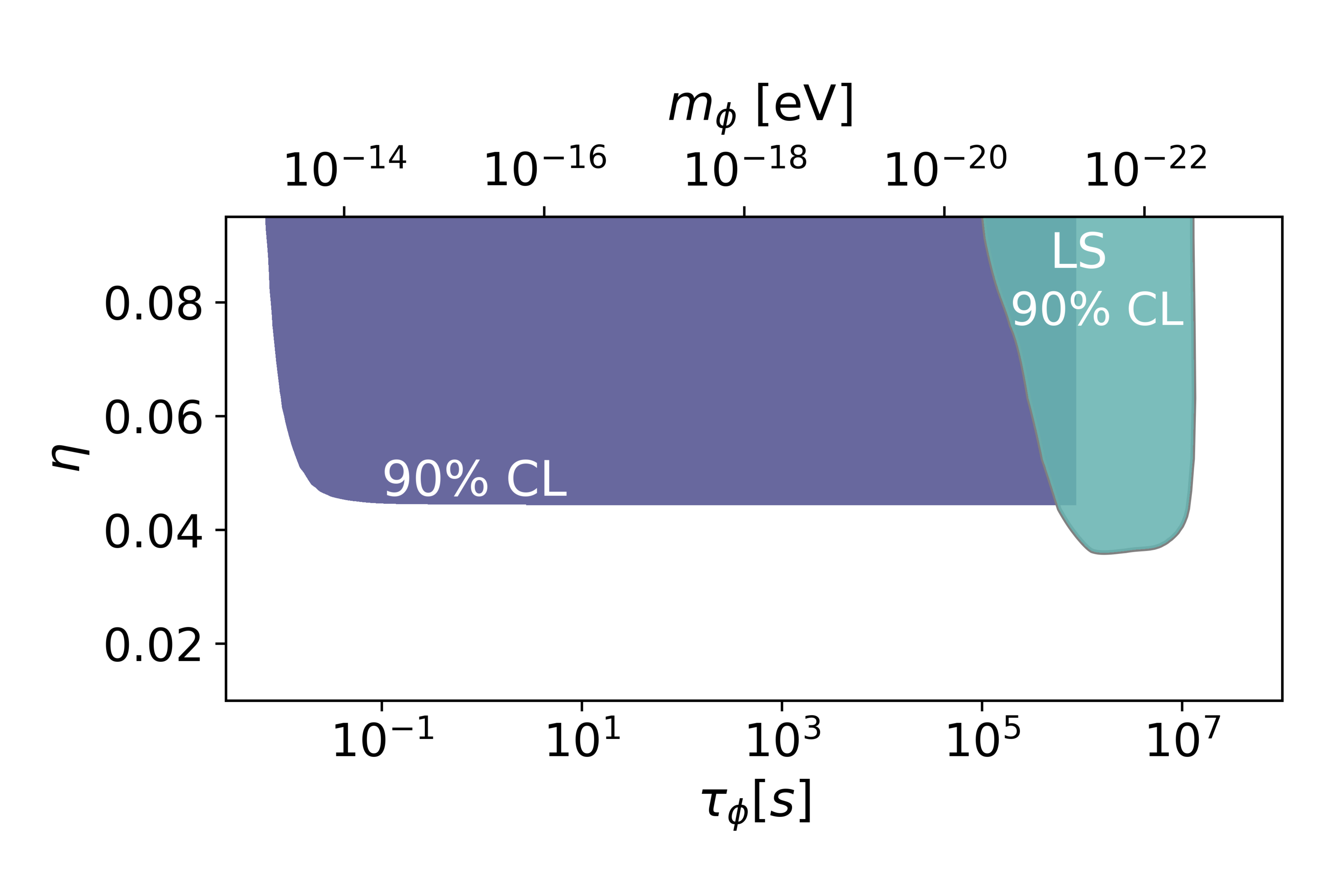}
    \caption{DUNE sensitivity (90\% C.L.) to ultralight scalars  via modulation of the atmospheric mass splitting $\Delta m^2_{31}$. The amplitude of modulation is $\eta$ (see Eqs.~\eqref{eq:angles} and \eqref{eq:dmsq}), and the modulation period and mass of the scalar field is denoted by $\tau_\phi$ and $m_\phi$, respectively.}
    \label{fig:all}
\end{figure}

Finally, we present in Fig.~\ref{fig:all}  DUNE's sensitivity to ultralight scalar dark matter incorporating the three searches proposed in this paper.
It is remarkable that a neutrino oscillation experiment can probe scalar masses ranging about 10 orders of magnitude. 
The transition between dynamical DiNOs and average DiNOs can be seen to be smooth.
When time modulation can be seen (LS labeled region), it provides an even better probe of ultralight scalars.
While its sensitivity spans a very large region in parameter space, DUNE would only be able to measure the mass of $\phi$ by the determination of the modulation frequency, which is only possible in the ``LS'' region.
Elsewhere, even if distorted neutrino oscillations are observed, it is not clear on how to determine the ultralight scalar mass.

\section{Other constraints on ultralight scalars}
Besides neutrino oscillations, ultralight bosonic dark matter may also be probed by other experiments and cosmological observations.
Several other constraints can be found in previous literature.
Here we simply summarize and comment on them.
\begin{itemize}
   \item CMB: As pointed out in previous work, the cosmological $\phi$ density redshifts as nonrelativistic matter, and thus the amplitude of modulation of $\phi$ is expected to be much larger at early times. This could lead to an increase in the sum of neutrino masses in the early universe, which would be constrained by observations of the Planck satellite~\cite{Ade:2015xua}. The constraint on the amplitude of $\phi$ is found to be $\eta\lesssim 9\times 10^{-3}$ if the atmospheric mass splitting modulates or $\eta \lesssim0.1$ if only the solar splitting modulates~\cite{Berlin:2016woy, Krnjaic:2017zlz, Brdar:2017kbt}. Note however that this constraint is model dependent. For example, if neutrino masses arrive from a seesaw mechanism and $\phi$ couples to, e.g., right-handed neutrinos, a large mass of the latter would actually imply a smaller mass of active neutrinos in the early universe.
   \item BBN: Following a similar reasoning, if $\phi$ couples universally to all standard model fermions, the changes in fermion masses could lead to observable effects in big bang nucleosynthesis, particularly on the abundance of $^4$He~\cite{Sibiryakov:2020eir}. In our scenario, we are only coupling $\phi$ to neutrinos, and therefore this bound would not apply. In a UV complete realization, this bound should be taken into account carefully.
   \item Astrophysical neutrinos and supernova 1987A: The coupling between neutrinos and the ultralight scalar could lead to scattering of astrophysical neutrinos on the dark matter background, making the universe opaque to certain astrophysical neutrinos. In particular, the observation of neutrinos from the supernova 1987A requires the universe to be transparent to MeV neutrinos. These constraints are several orders of magnitude weaker than the ones derived from neutrino oscillation measurements, except for large masses $m_\phi\gtrsim 10^{-8}$~eV where neutrino experiments lose sensitivity~\cite{Krnjaic:2017zlz, Brdar:2017kbt}.
   \item Electron mass modulation: Even if we postulate that $\phi$ only couples to neutrinos at tree level, a loop level coupling to electrons cannot be avoided, which could potentially lead to stronger constraints. Nevertheless, the magnitude of the loop-induced coupling is suppressed by $G_F m_e m_\nu$, too small to be observed~\cite{Krnjaic:2017zlz}.
   \item Black hole spin measurements: Ultralight bosons can be emitted copiously by rotating black holes, thereby suppressing their energy and angular momentum. The constraints from black hole spin measurements are typically in the parameter space within the region $10^{-21}<m_\phi<10^{-16}~$eV.
   Note that these bounds present some dependence on the systematic uncertainties related to black hole mass and spin observations, as well as
   on the details of the UV model (for instance, scalar self-interactions may weaken those bounds)~\cite{Baryakhtar:2017ngi, Davoudiasl:2019nlo, Zu:2020whs}.
\end{itemize}

\section{Conclusion}
In this paper, we have investigated in detail the phenomenology of ultralight bosonic dark matter fields in neutrino oscillation experiments.
We describe the three phenomenological regimes associated with different masses of the bosonic field, together with their respective signatures at neutrino oscillation experiments.
As a case study, we estimate the sensitivity of the DUNE experiment to ultralight scalars via detailed simulations and we provide a description of how to implement an experimental search for all regimes.
Finally, we show that the DUNE experiment is sensitive to about 10 orders of magnitude in ultralight scalar mass. 
The searches presented in this paper are general and can be easily applied to any other oscillation experiment.
Current experiments like MINOS, NOvA, T2K, Daya Bay, and RENO may perform the very first experimental searches for these scenarios.
In the future, besides DUNE, we also expect the JUNO to be particularly sensitive to modulations of mass splittings due to ultralight scalars.
Our results exhibit an excellent, unique opportunity to expand the physics program of neutrino oscillation experiments.

\section{Acknowledgements}
We thank Vedran Brdar, Zackaria Chacko, Gordan Krnjaic, Joachim Kopp, Alex Sousa and Jaehoon Yu for discussions. 
PMM is grateful for the kind hospitality of Fermilab theory group during the development of this work.
This manuscript has been authored by Fermi Research Alliance, LLC under Contract No. DE-AC02-07CH11359 with the U.S. Department of Energy, Office of Science, Office of High Energy Physics. PMM is supported by the FPU  grant  FPU18/04571, the Spanish grants FPA2017-90566-REDC (Red Consolider MultiDark), FPA2017-85216-P and SEV-2014-0398(MINECO/AEI/FEDER,  UE),  as  well  as  PROMETEO/2018/165  (Generalitat  Valenciana).  PMM received  partial  support  from  the  EU  Horizon 2020 project InvisiblesPlus (690575-InvisiblesPlus-H2020-MSCA-RISE-2015).   AD is supported in part by the National Science Foundation under Grant Number PHY-1914731.

\bibliography{bibliography}

\end{document}